\PassOptionsToPackage{tbtags,fleqn}{amsmath}
\documentclass[a4paper,UKenglish,cleveref,autoref,thm-restate]{lipics-v2021}
\title{Parallel Algorithms for the One Sided Crossing Minimization Problem} %TODO Please add

\author{Bogdan-Ioan Popa}{University of Bucharest, Faculty of Mathematics and Computer Science, Romania}{popabogdanpopa@gmail.com}{}{}%TODO mandatory, please use full name; only 1 author per \author macro; first two parameters are mandatory, other parameters can be empty. Please provide at least the name of the affiliation and the country. The full address is optional. Use additional curly braces to indicate the correct name splitting when the last name consists of multiple name parts.

\author{Adrian-Marius Dumitran}{University of Bucharest, Faculty of Mathematics and Computer Science}{marius.dumitran@unibuc.ro}{}{}

\author{Livia Magureanu}{University of Bucharest, Faculty of Mathematics and Computer Science}{livia.magureanu@gmail.com}{}{}

\authorrunning{B. Popa, M. Dumitran, L. Magureanu} %TODO mandatory. First: Use abbreviated first/middle names. Second (only in severe cases): Use first author plus 'et al.'

\Copyright{Bogdan-Ioan Popa, Adrian-Marius Dumitran and Livia Măgureanu} %TODO mandatory, please use full first names. LIPIcs license is "CC-BY";  http://creativecommons.org/licenses/by/3.0/

 %TODO mandatory: Please choose ACM 2012 classifications from https://dl.acm.org/ccs/ccs_flat.cfm 

 %TODO mandatory; please add comma-separated list of keywords

\category{} %optional, e.g. invited paper

\relatedversion{} %optional, e.g. full version hosted on arXiv, HAL, or other respository/website
\nolinenumbers
\begin{document}

\maketitle

\begin{abstract}
The One Sided Crossing Minimization (OSCM) problem is an optimization problem in graph drawing that aims to minimize the number of edge crossings in bipartite graph layouts. It has practical applications in areas such as network visualization and VLSI (Very Large Scale Integration) design, where reducing edge crossings improves the arrangement of circuit components and their interconnections. Despite the rise of multi-core systems, the parallelization of exact and fixed-parameter tractable (FPT) algorithms for OSCM remains largely unexplored. Parallel variants offer significant potential for scaling to larger graphs but require careful handling of synchronization and memory management. In this paper, we explore various previously studied exact and FPT algorithms for OSCM, implementing and analyzing them in both sequential and parallel forms. Our main contribution lies in empirically proving that these algorithms can achieve close to linear speedup under parallelization. In particular, our best result achieves a speedup of nearly 19 on a 16-core, 32-thread machine. We further investigate and discuss the reasons why linear speedup is not always attained.

\end{abstract}

\section{Related Work}

The One-Sided Crossing Minimization (OSCM) problem has been widely studied both theoretically and practically. Our work extends the parameterized complexity perspective in graph drawing and, to our knowledge, is the first to address OSCM in parallel computing.

\subsubsection*{Exact and Parameterized Algorithms}
Kobayashi and Tamaki achieved a major breakthrough with a subexponential FPT algorithm of $O(k2^{\sqrt{2k}}+n)$ \cite{kobayashi2015fast}, improving over the $O(1.4656^k+kn^2)$ algorithm of Dujmović, Fernau, and Kaufmann \cite{dujmovic2008fixed}. Their approach relied on interval graphs and dynamic programming. Similar advances have been made for related problems such as one-page (or circular) crossing minimization \cite{bannister2014parameterized}, showing strong theoretical progress for small $k$s.

\subsubsection*{Heuristics and Practical Approaches}
Due to the NP-hard nature of OSCM, many heuristics have been developed for practical applications. The barycenter and median heuristics from the Sugiyama framework remain widely used \cite{sugiyama1981methods, eades1994edge} methods due to their simplicity and effectiveness. Over the years, more advanced techniques have been explored, including stochastic methods, evolutionary algorithms, and local search (Di Battista et al. \cite{battista1998graph}). While these methods often provide high-quality solutions for larger instances, they lack optimality guarantees.

\subsubsection*{Distributed and Parallel Computing Perspectives}
Research on OSCM has focused on sequential algorithms, despite large-scale applications where sequential costs are prohibitive. Kobayashi and Tamaki’s interval graph structure \cite{kobayashi2015fast} suggests potential for parallelization, supported by existing parallel algorithms for interval graphs \cite{olariu1991optimal} and general frameworks for distributed dynamic programming \cite{bertsekas1989parallel}. Our work is the first to synthesize these concepts into a parallel algorithm specifically for OSCM.

\section{Introduction} 
Layered graph drawing is a prevalent method for visualizing hierarchical structures, widely known as the Sugiyama framework \cite{sugiyama1981methods}. A crucial step is crossing minimization, which aims to reduce the number of edge intersections between adjacent layers to enhance readability. The One-Sided Crossing Minimization (OSCM) problem fixes the vertex order on one layer and finds the optimal ordering for the vertices on the adjacent layer.

OSCM is NP-hard, even for sparse graphs \cite{eades1994edge, munoz2002one}, motivating research into various algorithmic approaches (heuristics, approximation algorithms, and fixed-parameter algorithms). An algorithm is fixed-parameter tractable (FPT) with respect to a parameter $k$ if it runs in $f(k) \cdot n^{O(1)}$ time. For OSCM, the number of crossings $k$ is a natural parameter.

A significant advance was the subexponential FPT algorithm by Kobayashi and Tamaki \cite{kobayashi2015fast}, which runs in $O(k2^{\sqrt{2k}} + n)$ time. Their elegant approach reduces OSCM to an interval graph problem and applies a dynamic programming solution. This algorithm made exact solutions feasible for instances where $k$ is not excessively large.

However, the sequential nature of this algorithm hinders its applicability on massive graphs such as social networks, protein-interaction networks, or large software architectures. This creates a pressing need for parallel algorithms capable of handling such large-scale instances.

\subsection{Contributions}

This work provides a systematic study of parallelization for exact and fixed-parameter algorithms addressing the One-Sided Crossing Minimization (OSCM) problem. Our main contributions are:

\begin{itemize}
    \item \textbf{Empirical parallelization study:} We implement and empirically analyze several exact and FPT algorithms for OSCM in both sequential and parallel settings.
    \item \textbf{Scalability evaluation:} We demonstrate that parallelization can yield close to linear speedup, with our best implementation achieving nearly 19× speedup on a 16-core, 32-thread machine.
    \item \textbf{Performance characterization:} We identify the cases where linear speedup is not attained and analyze the underlying causes.
    \item \textbf{Parallelization challenges:} We systematically discuss and address the key obstacles, including synchronization overhead, memory management and contention, load balancing across processing units.
    \item \textbf{Building a test environment:} We gathered and enhanced datasets to benchmark our implementations. We focus on specific types of graphs for each approach, depending on its weaknesses and strengths. 
\end{itemize}

\subsection{Preliminaries}

We are given a bipartite graph $G = (A, B, E)$ where $A = \{0, 1, \dots, n - 1\}$ and $B = \{0, 1, \dots, m - 1\}$ are the set of vertices and $E \subseteq A \times B$ the set of edges. Imagine drawing the vertices of $G$ on two straight parallel lines, vertices from $B$ are drawn in order from left to right, while vertices of $A$ can be drawn in any order we choose. The OSCM (one sided crossing minimization) problem asks to find an order in which to draw vertices from $A$ such that the total number of edge crossings is minimized.

More formally we are asked to find a permutation $P$ of $A$ such that the number of edge crossings relative to $P$ is minimized. We say that two edges $(a, b) \in E$ and $(c, d) \in E$ form a crossing relative to $P$ iff $P^{-1}_a < P^{-1}_c$ and $b > d$. Note that $P^{-1}_a$ is the position of vertex $a$ in permutation $P$.

We establish some notations which we will use throughout the paper:

For $x, y \in A$, let $C_{xy}$ be the number of edge intersections that vertices $x$ and $y$ create if $P^{-1}_x < P^{-1}_y$ (i.e. if $x$ is before $y$ in $P$). By definition $C_{xx} = 0$.

Then we define $F(Y, x) = \sum_{y \in Y} C_{yx}$ for $x \in A$ and $Y \subseteq A$.

We use $N_a = \{b \mid (a, b) \in E\}$ to denote the set of neighbours of node $a$, and $d_a = |N_a|$ which is the degree of node $a$.

For an algorithm, let $T_p$ be the time it takes for the algorithm to finish executing while running on $p$ processors. The speedup is then calculated by dividing the time it takes to finish with one processor by the time it takes with $p$ processors $S_p = \frac{T_1}{T_p}$.

The algorithms were implmeneted in \textit{C++20} and tested under Ubuntu 24.04 LTS on a \textit{32GB} of RAM machine with a Ryzen 9 5950X running at 4.9GHz, with 16 cores and 32 threads.

% \subsubsection{The Interval Graph Formulation}
% The algorithm by Kobayashi and Tamaki \cite{kobayashi2015fast} is based on a key insight. For each vertex $y \in Y$, let $l_y$ and $r_y$ be the smallest and largest neighbors of $y$ in $X$ with respect to the fixed order $<$. We can associate each vertex $y \in Y$ with a half-open interval $I_y = [l_y, r_y)$.

% A crucial observation is that if two intervals $I_u$ and $I_v$ do not intersect (e.g., $r_u \le l_v$), then the relative ordering of $u$ and $v$ is fixed in any optimal solution. Specifically, if $r_u \le l_v$, then $u$ must precede $v$ to minimize crossings. The pairs of vertices $\{u, v\}$ whose relative ordering is not fixed are precisely those for which the intervals $I_u$ and $I_v$ intersect. These are called \emph{orientable pairs}.

% This transforms the problem into finding an optimal orientation for the edges of an \emph{interval graph} $\mathcal{G_I} = (Y, E_I)$, where an edge $(u,v) \in E_I$ exists if and only if intervals $I_u$ and $I_v$ intersect. The algorithm then uses dynamic programming on a path decomposition of this interval graph to find the optimal ordering. The width of this path decomposition is bounded by the size of the maximum clique in $\mathcal{G_I}$, which for a feasible instance is at most $\sqrt{2k}+1$. This bound is key to the algorithm's subexponential runtime.

In what follows, we present several algorithms for the OSCM problem along with our implementations, including both sequential and parallel versions. We provide benchmarking results for these algorithms and analyze the impact and effectiveness of parallelization under various conditions.

\section {Bitmask DP Algorithms} \label{bitmask-dp-approaches}

We propose several variants of an algorithm for solving the OSCM problem based on a classical dynamic programming approach, which have different time and memory complexities. We implement the sequential and parallel versions of each variant, and then test them head to head on the same dataset to measure their execution times and speedups.

\subsection{High-Level Algorithm Overview}

Let $X \subseteq A$ be a subset of vertices from $A$, then:

\begin{tabular}{rl}
    $dp_X = $ & the minimum number of crossings we can obtain by considering only the vertices of \\
     & the subset $X$.
\end{tabular}

It is obvious to see that $dp_{\emptyset} = 0$ and the solution to the problem is found in $dp_A$. The recurrence for this dynamic programming approach follows:

$$dp_X = \min_{v \in X} dp_{X \setminus \{v\}} + F(X \setminus \{v\}, v)$$

% The recurrence comes from iterating through the set of vertices to find the best vertex to be placed last in the permutation of $X$ which minimizes the crossings. Let $v$ be such a vertex. The rest of the vertices can be optimally permuted to get $dp_{X \setminus \{v\}}$ crossings, and adding $v$ to the end of the permutation yields an extra $F(X \setminus \{v\}, v)$ crossings. Having a subset $X$ we can iterate through vertices $v$ of X and find which one gave us the optimal value for $dp_X$. 
See a basic implementation of this approach in [\ref{appendix:bitmask-template}].

The great thing about this algorithm is that, with minimal modifications, it can be parallelized to achieve linear speedup. Ideally, we want to split the work of computing the $dp$ array evenly between our processors. However, splitting the range $[1, \dots, 2^n - 1]$ evenly into intervals and assigning an interval to each processor won't work, as a processor will require the value $dp_Y$ of some $Y$ which may not have been calculated yet. To solve this problem we can choose another order of calculating the $dp$ array. If $X$ has $k$ bits set, then the value of $dp_X$ depends on the values of $dp_Y$, where $Y$ has $k - 1$ bits set. Thus, we can distribute the states into $n$ layers, the $i^{th}$ layer having assigned the states which have $i$ bits set, and parallelize the calculation of states on the same layer. See a sequential implementation that computes the $dp$ array state by state in the order of the layers in \ref{appendix:bitmask-sequential}.

To parallelize the DP algorithm, we will compute the $dp$ values in the same layer in parallel. For each available processor we can assign its work as $work_j = \ceil{\frac{{n \choose i}}{PROC}}$ if $j < {n \choose i} \mod PROC$, else $work_j = \floor{\frac{{n \choose i}}{PROC}}$, where $PROC$ is the number of available processors. The first processor takes the first $work_0$ states, the second processor takes the next $work_1$ states, and so on. See a parallel implementation that computes the $dp$ array layer by layer in \ref{appendix:bitmask-parallel}.

Depending on how we compute the $F$ function, our time complexity can vary from $O(2^nn^2 + n^2m)$ to $O(2^nn + n^2m)$. The slow version chooses a naive approach, iterating through all $y$'s and summing $C_{yx}$, thus adding an $n$ factor to the time complexity. The fast versions precompute the $F$ function using $O(2^nn)$ and $O(2^{n / 2} n)$ additional memory respectively.

\subsection{Slow Bitmask DP}

The slow version of the bitmask DP implements the $F$ function in $O(n)$, thus the resulting algorithm runs in $O(2^nn^2 + n^2m)$. Due to its relatively small memory consumption, its implementation only requires storing in memory the graph, the $C$ matrix and the $dp$ array. This allowed us to test this implementation with $n$ up to $30$, resulting the following running stats:

\begin{figure}[H]
\centering
\caption{Running stats for slow\_bitmask\_dp}

\includegraphics[scale=0.5]{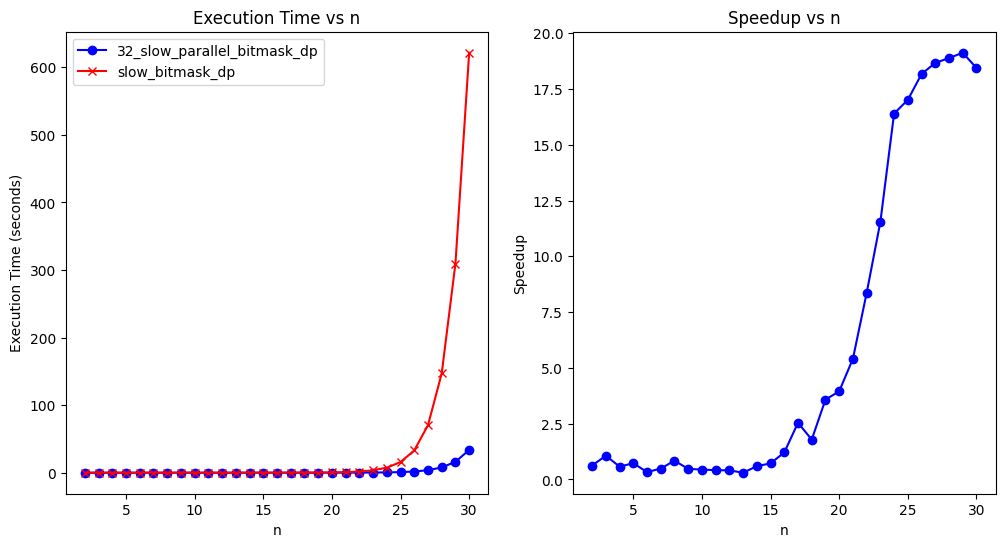}
\end{figure}

\begin{table}[h!]
\centering
\caption{Comparison of performance counter statistics: \texttt{slow\_bitmask\_dp} sequential vs parallel execution}
\label{bitmask:slow-dp-table}
\begin{tabular}{|l|r|r|l|}
\hline
\textbf{Metric} & \textbf{Sequential} & \textbf{Parallel} & \textbf{Notes} \\
\hline
Task-clock (msec)           & 745,218.44     & 1,297,914.14   & CPUs utilized: 1 vs 29.652 \\
Context-switches            & 1,038          & 2,571          & /sec: 1.39 vs 1.98 \\
CPU-migrations              & 0              & 12             & negligible \\
Page-faults                 & 1,048,717      & 1,050,414      & nearly identical \\
Cycles                      & 2433601012816  & 3291638160915          & 3.266 GHz vs 2.536 GHz \\
Instructions                & 3023756184522 & 3291440594184          & IPC: 1.24 vs 1.00 \\
Branches                    & 868386509436  & 943469518286        & rate: 1.165 G/sec vs 727 M/sec \\
Branch-misses               & 89782828418 & 79178528101 & 10.34\% vs 8.39\% of branches \\
Cache-references            & 11797983262          & 22238880018          & per sec: 15.8M vs 17.1M \\
Cache-misses                & 334257425 & 1465155185 & 2.83\% vs 6.59\%  of cache refs\\
L1-dcache-loads             & 525730337123         & 551102795385         & per sec: 705M vs 425M \\
L1-dcache-load-misses       & 9001114451  & 13958309953 & 1.71\% vs 2.53\% of L1 accesses \\
\hline
\end{tabular}
\end{table}

The parallel version achieved a $\approx$ 19× speedup compared to the sequential run. The high speedup is explained by excellent parallel scalability: the workload distributes well across threads, and the memory behavior remains relatively efficient (low cache-miss rates, stable IPC). The parallel run spends more aggregate CPU time (higher task-clock), but the wall-clock time seen by the user drops significantly because 30 cores are effectively utilised. See \autoref{bitmask:slow-dp-table}.

\subsection{Fast Bitmask DP}

This version of the algorithm reduces the complexity of the $F$ function to $O(1)$ by precomputing it using the following recurrence relation:

$$F(Y, x) = F(Y \setminus \{y\}, x) + C_{yx} \text{ for some } y \in Y$$

\begin{figure}[H]
\centering
\caption{Running stats for fast bitmask DP with high memory}
\includegraphics[scale=0.5]{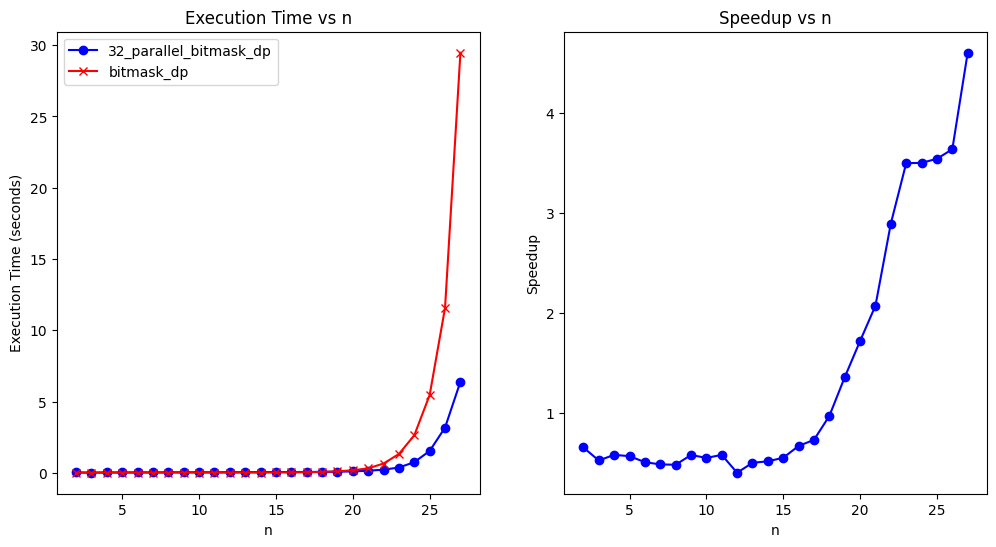}
\end{figure}

This precomputation requires storing a big $2-$dimensional array in memory of size $2^nn$, making our \texttt{C++} implementation only possible to be tested on inputs up with $n$ up to $27$, due to hardware limits (\ref{appendix:fast-bitmask}). This version actually performs worse in practice than the low-memory one due to its high memory consumption and expensive computation.

The precomputation of the $F$ function also takes $O(2^nn)$, so the asymptotic time complexity of our algorithm does not change. Because this is also an expensive computation, which has the same running time as calculating the $dp$ array, it also has to be parallelized. To parallelize this part of the algorithm we can just run the first for loop in parallel. 

\begin{table}[h!]
\centering
\caption{Comparison of performance counter statistics: \texttt{bitmask\_dp} sequential vs parallel execution}
\label{bitmask:fast-dp-table}
\begin{tabular}{|l|r|r|l|}
\hline
\textbf{Metric} & \textbf{Sequential} & \textbf{Parallel} & \textbf{Notes} \\
\hline
Task-clock (msec)           & 35,288.04  & 174,029.07 & CPUs utilized: 1 vs 23.069 \\
Context-switches            & 102        & 940        & /sec: 2.890 vs 5.401 \\
CPU-migrations              & 0          & 11         & /sec: 0.000 vs 0.063 \\
Page-faults                 & 5,111,949  & 3,671,815  & /sec: 144.863 K vs 21.099 K \\
Cycles                       & 111,242,915,404 & 477,378,956,856 & 3.152 GHz vs 2.743 GHz (util \%) \\
Instructions                 & 123,789,150,846 & 120,975,873,317 & insn per cycle: 1.11 vs 0.25 \\
Branches                     & 26,642,198,104 & 23,756,601,207 & M/sec: 754.992 vs 136.509 \\
Branch-misses                & 1,454,284,994  & 1,282,291,082  & 5.46\% vs 5.40\% of branches \\
Cache-references             & 7,655,945,356  & 4,628,036,668  & M/sec: 216.956 vs 26.593 \\
Cache-misses                 & 1,222,024,085  & 1,162,552,160  & 15.96\% vs 25.12\% of cache refs \\
L1-dcache-loads              & 46,441,160,046 & 32,085,031,330 & G/sec: 1.316 vs 0.184 \\
L1-dcache-load-misses        & 4,043,146,576  & 3,341,266,421  & 8.71\% vs 10.41\% of L1 accesses \\
\hline
\end{tabular}
\end{table}

The parallel version achieved only a $\approx$ 4.7× speedup on $\approx$ 23 cores. Here, the counters show a significant drop in instructions per cycle (IPC) (from 1.11 to 0.25) and an increase in cache-miss ratios (cache-miss \% and L1-miss \% both worsen). This indicates that threads compete heavily for memory bandwidth and shared caches, causing pipeline stalls. Although the total number of instructions executed is similar to the sequential run, per-core throughput drops sharply due to memory contention. Consequently, adding threads provides only modest gains, as execution becomes memory-bound rather than compute-bound. See \autoref{bitmask:fast-dp-table}.

\subsection{Low-memory Fast Bitmask DP}

We can reduce the memory usage of the algorithm to $O(2^n)$ while preserving the time complexity of $O(2^nn)$ by using a technique called \textit{"Meet in the Middle"}. We can split the set of vertices $A$ into $A_1 = \{0, \dots, \floor{\frac{n}{2}} - 1\}, A_2 = \{\floor{\frac{n}{2}}, \dots, n - 1\}$, then precompute: $F_1(Y, x) = \sum_{y \in Y} C_{yx}$, where $Y \subseteq A_1$, and $F_2(Y, x) = \sum_{y \in Y} C_{yx}$, where $Y \subseteq A_2$. Thus, $F(Y, x) = F_1(Y \cap A_1, x) + F_2(Y \cap A_2, x)$. This idea drastically reduces not only the precomputation time of the $F$ function to $O(2^{\frac{n}{2}}n)$, but also the memory complexity. Please refer to \autoref{appendix:low-memory-bitmask} for implementation details.

\begin{figure}[H]
\centering
\caption{Running stats for fast bitmask DP with low memory}
\includegraphics[scale=0.5]{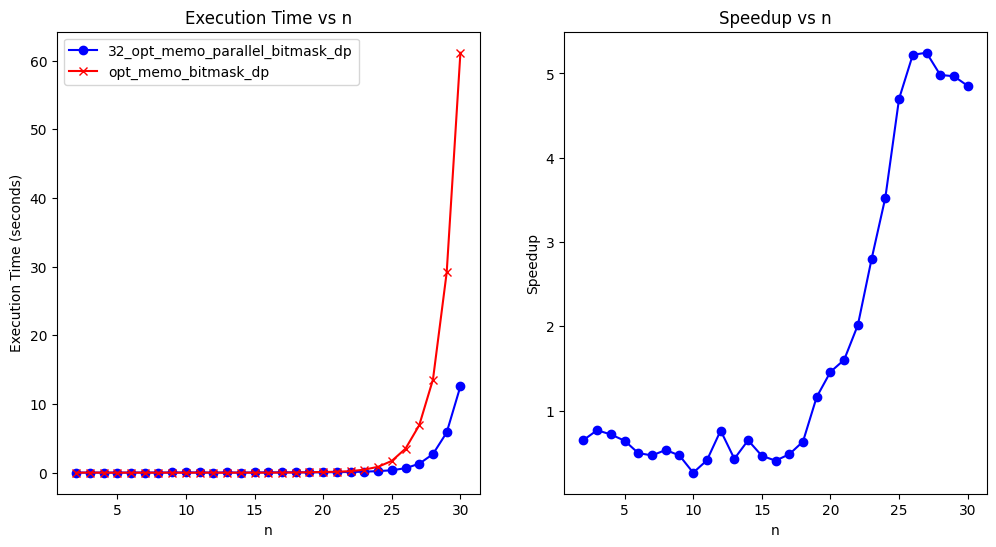}
\end{figure}

\begin{table}[h]
\centering
\caption{Comparison of performance counter statistics: \texttt{opt\_memo\_bitmask\_dp} sequential vs parallel execution}
\label{bitmask:low-memory-table}
\begin{tabular}{|l|r|r|l|}
\hline
\textbf{Metric} & \textbf{Sequential} & \textbf{Parallel} & \textbf{Notes} \\
\hline
Task-clock (msec)           & 70,572.73  & 377,468.70 & CPUs utilized: 1 vs 23.720 \\
Context-switches            & 98         & 1,563      & /sec: 1.389 vs 4.141 \\
CPU-migrations              & 0          & 28         & /sec: 0.000 vs 0.074 \\
Page-faults                 & 1,051,151  & 1,052,533  & /sec: 14.895 K vs 2.788 K \\
Cycles                       & 228,653,344,897 & 993,895,629,335 & 3.240 GHz vs 2.633 GHz (util \%) \\
Instructions                 & 396,993,999,185 & 717,875,113,483 & insn per cycle: 1.74 vs 0.72 \\
Branches                     & 69,602,327,320  & 149,291,619,046 & M/sec: 986.250 vs 395.507 \\
Branch-misses                & 6,608,992,320   & 6,482,413,460   & 9.50\% vs 4.34\% of branches \\
Cache-references             & 25,747,201,061  & 41,278,889,966  & M/sec: 364.832 vs 109.357 \\
Cache-misses                 & 1,590,200,248   & 6,453,645,164   & 6.18\% vs 15.63\% of cache refs \\
L1-dcache-loads              & 120,191,927,259 & 182,429,050,206 & G/sec: 1.703 vs 0.483 \\
L1-dcache-load-misses        & 15,483,151,684  & 37,041,566,827  & 12.88\% vs 20.30\% of L1 accesses \\
\hline
\end{tabular}
\end{table}

The parallel version of \texttt{opt\_memo\_bitmask\_dp} shows a clear utilization of multiple cores, averaging 23.7 effective CPUs out of the 32 threads launched. This explains the significant reduction in wall-clock time compared to the sequential version. However, the efficiency is not perfect: the instructions-per-cycle (IPC) drops from 1.74 to 0.72, indicating that threads spend more time stalled, likely due to increased memory traffic and synchronization. Cache pressure is also evident: cache-misses rise from 6.18\% to 15.63\%, and L1-dcache-load-misses grow from 12.9\% to 20.3\%, which reflects contention and reduced locality in the parallel run. See \autoref{bitmask:low-memory-table}.

\subsection{Dataset generation}

Due to the algorithm's design, its running time is only marginally affected by the layout of the input graph, so we can randomly generate bipartite graphs to test its performance. The above implementations were tested on consecutive values of $n$, starting from $2$, until the running time increased too much or the testing computer ran out of RAM memory.

\section {Golden Ratio FPT}

In 2002, Dujmović and Whitesides gave an $O(\phi^kn^2 + n^2k)$ time FPT algorithm  \cite{golden_ratio}, where $\phi = \frac{1 + \sqrt{5}}{2}$, the golden ratio, and $k$ is the maximum number of tolerated crossings. In the following section, we will present its idea and share its implementation details, along with some benchmark results. We also give ideas on our take on parallelizing this algorithm and compare the benchmark results to the sequential version. 

\subsection{High-Level Algorithm Overview}

First, we compute $C_{xy}$ for every pair $x, y \in A$. The algorithm is based on a fairly common approach on FPT algorithms, that is Bounded Search Trees. Here, we build a search tree that is then exhaustively traversed to find a solution to the problem. We define an instance of the problem as a pair $I = (D, B)$ where $D$ is an $n \times n$ array where $D_{ij} = 1$ if $i$ appears before $j$ in the final permutation, and $B$ is $k$ minus the current number or crossings generated by the order imposed by $D$ so far.  Since we do not know the value of $k$, we will start with $k = 1$ and then double it until we find the optimal solution. For more details, please refer to the pseudocode implementation \autoref{appendix:golden-ratio-fpt} or \cite{golden_ratio}.

For our tests, we ran the implemented algorithm on tests with $n$ upwards of $16.000$. Our goal was to minimize the preprocessing time, such that we can truly see how much work the searching part of the algorithm does. Even though the transitive closure can be computed in $O(n^3)$ using Floyd-Warshall algorithm, we opted for a slightly better $O(\frac{n^3}{W})$ algorithm which runs better in practice where $W$ is the word size of the processor which runs the algorithm.

To speed up our implemented algorithm we performed some additional optimizations. We know that $OPT \geq LB$, where $LB = \sum_{(i, j)}min(C_{ij}, C_{ji})$, so we only search for solutions with crossing numbers $\leq LB + k$. Additionally, for each pair $(i, j)$ we compute $C'_{ij} = C_{ij} - min(C_{ij}, C_{ji})$. Check the final implementations in \ref{appendix:golden-ratio-tr-closure} and \ref{appendix:golden-ratio-search-function}.

\subsection{Paralellizing the algorithm}

To parallelize the above algorithm we used a thread pool. Let $PROC$ be the number of available processors. We spawn $PROC$ threads that search through the tree concurrently. The thread pool has a stack that all the threads share, where instances to be explored are pushed. Threads then wait to extract instances from the stack in a synchronized manner and insert new child instances into the stack. 

The thread pool is implemented as a class with the following members:
% \begin{itemize}
%     \item \textit{threads}, an array of $PROC$ threads.
%     \item \textit{S}, a stack of instances that will be explored.
%     \item \textit{working}, a variable that indicates how many threads are executing a search call
%     \item \textit{mtx}, a mutex for synchronized access to \textit{S} and \textit{working}
%     \item \textit{cv}, a condition variable which helps notify threads if any events take place, such as a new search call has been pushed to the stack or the stack is empty and no more threads are working.
% \end{itemize}

\begin{tabular}{rcl}
    \textit{threads} & $\rightarrow$ & an array of $PROC$ threads \\
    \textit{working} & $\rightarrow$ & a variable that indicates how many threads are executing a search call \\
    \textit{S} & $\rightarrow$ & a stack of instances that will be explored \\
    \textit{mtx} & $\rightarrow$ & a mutex for synchronized access to \textit{S} and \textit{working} \\
    \textit{cv} & $\rightarrow$ & a condition variable which helps notify threads if any events take place, such \\
    & & as a new search call has been pushed to the stack or the stack is empty and \\
    & & no more threads are working
\end{tabular}
    
See \autoref{appendix:golden-ratio-parralel} for implementation details.

\subsection{Dataset and experiments}

To test our implementations, we used the tests from the exact track of the PACE 2024 \cite{pace2024} competition as well as our own generated tests. Due to the exponential nature of our algorithm, we filtered tests from the PACE 2024 \cite{pace2024} competition in order to keep the tests for which that took a reasonable time to finish and to discard the test cases where the algorithm would finish too quickly because we deemed them irrelevant for benchmarking purposes. 

\begin{figure}[h]
\centering
\caption{Running stats for the golden ratio FPT algorithm}
\includegraphics[scale=0.45]{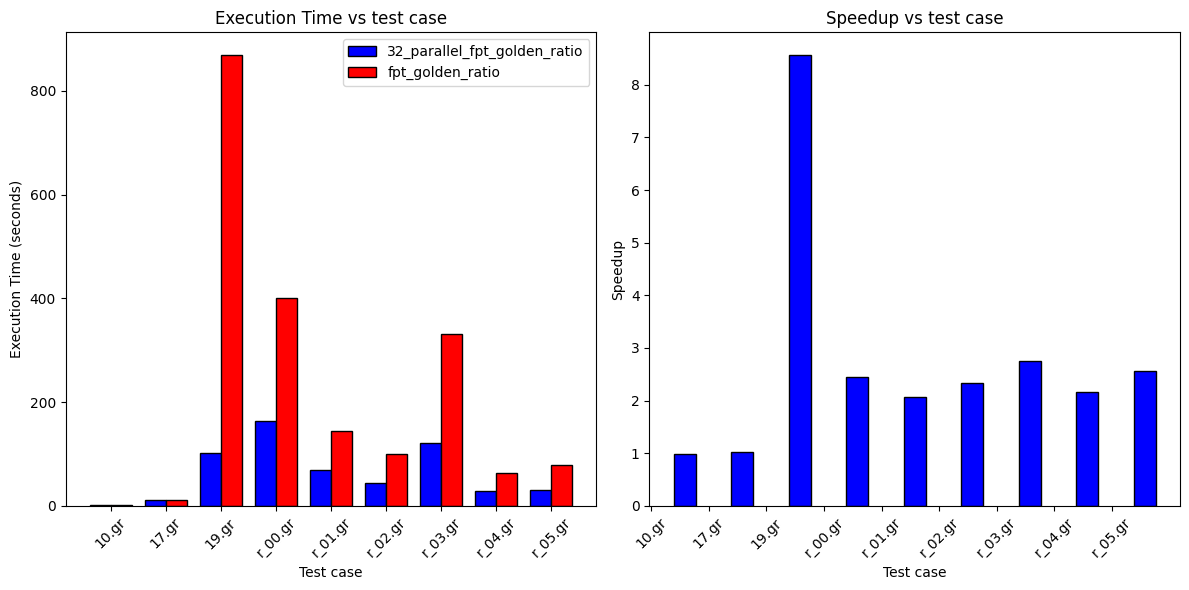}
\end{figure}

Our experiments show that depending on the test cases used, we can achieve different speedups. This comes from the structure of the search tree. If the search tree has lots of deep branches, then parallelizing the search indeed gives us significant gains. Otherwise, a parallel variant of the algorithm can even work against us, giving us slower times, because of the overhead of the threads waiting and battling over which one extracts from the stack.

\begin{table}[H]
\centering
\caption{Comparison of performance counter statistics: \texttt{fpt\_golden\_ratio} sequential vs parallel execution on test \texttt{19.gr}}
\begin{tabular}{|l|r|r|l|}
\hline
\textbf{Metric} & \textbf{Sequential} & \textbf{Parallel} & \textbf{Notes} \\
\hline
Task-clock (msec)           & 1,013,317.52 & 5,217,149.09 & CPUs utilized: 1 vs 31.769 \\
Context-switches            & 1,347        & 206,247      & /sec: 1.329 vs 39.533 \\
CPU-migrations              & 1            & 1,630        & /sec: 0.001 vs 0.312 \\
Page-faults                 & 31,035       & 939,717      & /sec: 30.627 vs 180.121 \\
Cycles                       & 3,260,761,955,076 & 12,610,396,539,561 & 3.218 GHz vs 2.417 GHz (util \%) \\
Instructions                 & 10,820,215,895,396 & 20,591,317,067,420 & insn per cycle: 3.32 vs 1.63 \\
Branches                     & 2,938,160,977,014  & 5,644,050,720,371  & G/sec: 2.900 vs 1.082 \\
Branch-misses                & 3,720,876,477      & 10,968,267,343     & 0.13\% vs 0.19\% of branches \\
Cache-references             & 163,473,562,499    & 467,379,523,217    & M/sec: 161.325 vs 89.585 \\
Cache-misses                 & 35,881,277,150     & 174,337,718,468    & 21.95\% vs 37.30\% of cache refs \\
L1-dcache-loads              & 2,086,256,316,193  & 4,122,099,058,614  & G/sec: 2.059 vs 0.790 \\
L1-dcache-load-misses        & 101,065,151,238    & 231,153,166,810    & 4.84\% vs 5.61\% of L1 accesses \\
\hline
\end{tabular}
\end{table}

On the large instance \texttt{19.gr}, the parallel version of \texttt{fpt\_golden\_ratio} effectively utilized an average of \textbf{31.8 CPUs}, close to the 32 threads launched, demonstrating strong scalability in terms of raw parallelism. This allowed the program to complete much faster in wall-clock time despite higher total work ($\approx$ 8x speedup). However, efficiency drops significantly: the \textbf{IPC} falls from 3.32 to 1.63, and both the \textbf{cache-miss rate} (21.95\% $\rightarrow$ 37.30\%) and the \textbf{L1-dcache-load-miss rate} (4.84\% $\rightarrow$ 5.61\%) increase. This indicates that the parallel threads interfere in the memory hierarchy, stressing caches and lowering per-core throughput. 

\begin{table}[h!]
\centering
\caption{Comparison of performance counter statistics: \texttt{fpt\_golden\_ratio} sequential vs parallel execution on test \texttt{r\_01.gr}}
\begin{tabular}{|l|r|r|l|}
\hline
\textbf{Metric} & \textbf{Sequential} & \textbf{Parallel} & \textbf{Notes} \\
\hline
Task-clock (msec)           & 171,237.64   & 4,488,782.68 & CPUs utilized: 1 vs 29.012 \\
Context-switches            & 172          & 10,090,123   & /sec: 1.004 vs 2.248 K \\
CPU-migrations              & 0            & 17,179       & /sec: 0.000 vs 3.827 \\
Page-faults                 & 635          & 1,579        & /sec: 3.708 vs 0.352 \\
Cycles                       & 536,320,543,377  & 6,274,458,097,701  & 3.132 GHz vs 1.398 GHz (util \%) \\
Instructions                 & 2,240,521,258,513 & 3,723,337,901,941 & insn per cycle: 4.18 vs 0.59 \\
Branches                     & 407,414,144,085   & 716,091,810,490   & 2.379 G/sec vs 159.529 M/sec \\
Branch-misses                & 1,325,435,088     & 11,619,360,021    & 0.33\% vs 1.62\% of branches \\
Cache-references             & 2,905,054,489     & 72,845,445,959    & M/sec: 16.965 vs 16.228 \\
Cache-misses                 & 641,024,253       & 17,021,779,588    & 22.07\% vs 23.37\% of cache refs \\
L1-dcache-loads              & 477,649,893,118   & 985,969,208,838   & G/sec: 2.789 vs 0.220 \\
L1-dcache-load-misses        & 1,720,812,453     & 29,507,732,433    & 0.36\% vs 2.99\% of L1 accesses \\
\hline
\end{tabular}
\end{table}

On instance \texttt{r\_01.gr}, the parallel execution of \texttt{fpt\_golden\_ratio} utilized an average of \textbf{29 CPUs} out of 32, which shows high parallel occupancy. However, the sequential version achieved a very high \textbf{IPC} of 4.18, while the parallel version dropped to only 0.59, indicating substantial memory stalls and synchronization overheads. The nearly constant cache reference rate (16--17 M/sec) suggests the bottleneck is not the number of memory accesses but the increased latency due to interference across threads.

\section {Subexponential FPT}

In 2014, Yasuaki Kobayashi and Hisao Tamaki gave a subexponential $O(k2^{\sqrt{2k}} + n)$ time fixed parameter algorithm \cite{fpt_subexpo}, where $k$ is the maximum number of tolerable edge crossings. It is a dynamic programming algorithm similar in nature with the bitmask dp algorithm presented in \autoref{bitmask-dp-approaches}. This algorithm exploits an interval graph associated with each OSCM instance, which was not fully explored in earlier work. In the following section, we will roughly present its idea and share its implementation details, along with some benchmark results. We also give ideas on our take on parallelizing this algorithm and compare the benchmark results with the sequential version. 

\subsection{High-Level Algorithm Overview}

We use the interval system  $J_a = [x_a, y_a]$, mentioned in \cite{fpt_subexpo} and maintain the set $M_t$ of intervals (vertices) that are active at point $t \in \{1, 2, \ldots, 2n\}.$

We compute $dp_Y = $ the minimum number of crossings considering only the vertices $Y \subseteq A$. Let $L_t = \{a \mid y_a \leq t\}$, then as shown in \cite{fpt_subexpo}, we can incrementally compute the dp array, at each step $t \in \{1, \dots, 2n\}$ we compute $dp_{L_t \bigcup S}$ for each $S \subseteq M_t$, because the vertex $v \in L_t \bigcup S$ that can be placed last in the permutation must be in $S$.

Computing $F(L_t \bigcup S \setminus \{v\}, v)$ can be done by first splitting $F(L_t \bigcup S \setminus \{v\}, v) = F(L_t, v) + F(S \setminus \{v\}, v)$. $F(L_t, v)$ can be maintained while traversing with t, and $F(S \setminus \{v\}, v)$ can be computed in $O(1)$ with an $O(2^{\frac{|M_t|}{2}}|M_t|)$ precomputation similar to \ref{appendix:low-memory-bitmask}. For more implementation details see \autoref{appendix:subexponential-fpt}.

\subsection{Paralellizing the algorithm}

Due to its similarity to the Bitmask DP approach, this algorithm is easily implemented in parallel. We split $S \subseteq M_t$ into $|M_t|$ layers. In layer $i$ we put $dp_{L_t \bigcup S}$ with $|S| = i$, then for each layer $i$, we distribute the $|M_t| \choose i$ states evenly to the $PROC$ available processors, then we compute each state similar to \ref{appendix:bitmask-parallel}.

\subsection{Dataset and experiments}

To test our implementations, we used the tests from the exact track of the PACE 2024 \cite{pace2024} competition as well as our own generated tests. We filtered tests which have the maximum value of $M_t \leq 30$, resulting in $11$ test cases.

\begin{figure}[h]
\centering
\caption{Running stats for the subexponential fpt algorithm}
\includegraphics[scale=0.5]{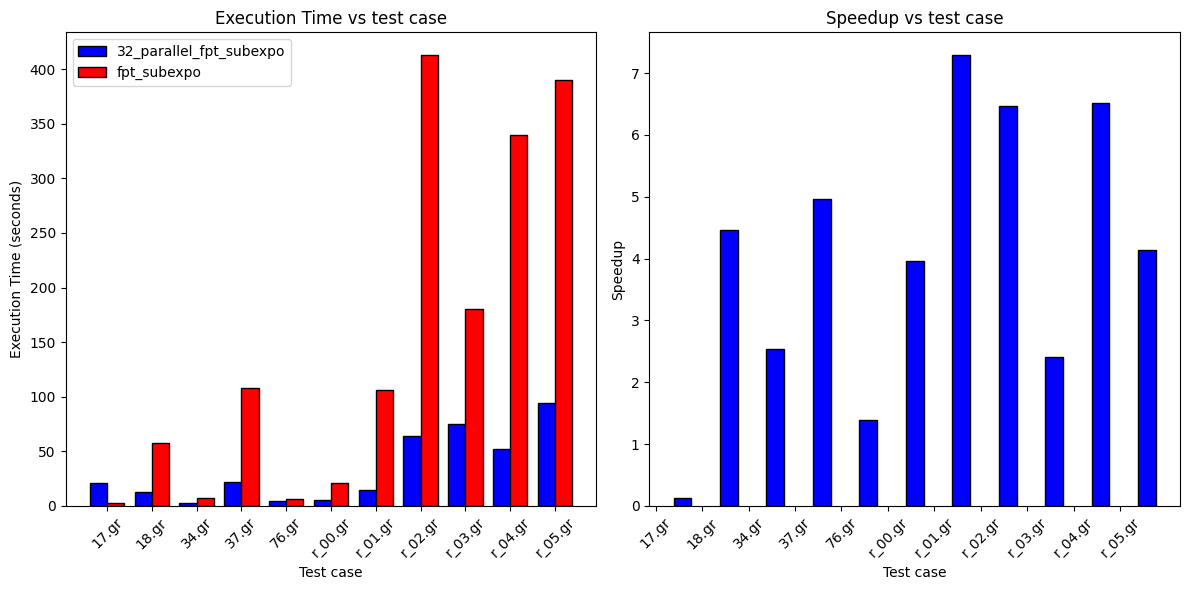}
\end{figure}

\begin{table}[h!]
\centering
\caption{Comparison of performance counter statistics: \texttt{fpt\_subexpo} sequential vs parallel execution on test \texttt{r\_01.gr}}
\label{subexponential:table_r_01.gr}
\begin{tabular}{|l|r|r|l|}
\hline
\textbf{Metric} & \textbf{Sequential} & \textbf{Parallel} & \textbf{Notes} \\
\hline
Task-clock (msec)           & 126,628.33   & 461,288.80   & CPUs utilized: 1 vs 18.901 \\
Context-switches            & 175          & 64,086       & /sec: 1.382 vs 138.928 \\
CPU-migrations              & 0            & 1,111        & /sec: 0.000 vs 2.408 \\
Page-faults                 & 200,233      & 328,490      & /sec: 1.581 K vs 712.114 \\
Cycles                       & 397,970,867,873  & 1,078,984,006,128 & 3.143 GHz vs 2.339 GHz (util \%) \\
Instructions                 & 633,822,927,397  & 1,080,615,444,876 & insn per cycle: 1.59 vs 1.00 \\
Branches                     & 108,693,044,509  & 233,351,693,147   & 858.363 M/sec vs 505.869 M/sec \\
Branch-misses                & 12,581,957,534   & 12,169,720,365    & 11.58\% vs 5.22\% of branches \\
Cache-references             & 31,718,927,508   & 64,537,919,140    & M/sec: 250.488 vs 139.908 \\
Cache-misses                 & 807,068,859      & 8,011,651,241     & 2.54\% vs 12.41\% of cache refs \\
L1-dcache-loads              & 201,062,723,351  & 244,615,832,195   & G/sec: 1.588 vs 0.530 \\
L1-dcache-load-misses        & 21,151,398,807   & 51,772,857,200    & 10.52\% vs 21.16\% of L1 accesses \\
\hline
\end{tabular}
\end{table}

On the \texttt{r\_01.gr} instance, the parallel \texttt{fpt\_subexpo} achieved an average of \textbf{18.9 CPUs}, demonstrating moderate parallelism. The per-core efficiency decreased compared to the sequential run: IPC dropped from 1.59 to 1.00, reflecting pipeline stalls and synchronization overhead. Memory performance also degraded, with a higher cache-miss rate (12.41\% vs 2.54\%) and increased L1-dcache-load misses (21.16\% vs 10.52\%), indicating cache contention.

Despite these inefficiencies, the parallel version benefits from large $|M_t|$ values for all $t \in \{1, 2, \ldots, 2n\}$, resulting in a net speedup over the sequential execution. See \autoref{subexponential:table_r_01.gr}.

\begin{table}[h!]
\centering
\caption{Comparison of performance counter statistics: \texttt{fpt\_subexpo} sequential vs parallel execution on testcase \texttt{17.gr}}
\label{subexponential:table_17.gr}
\begin{tabular}{|l|r|r|l|}
\hline
\textbf{Metric} & \textbf{Sequential} & \textbf{Parallel} & \textbf{Notes} \\
\hline
Task-clock (msec)           & 3,121.35      & 248,303.51    & CPUs utilized: 1 vs 2.420 \\
Context-switches            & 3             & 337,290       & /sec: 0.961 vs 1.358 K \\
CPU-migrations              & 0             & 3,284         & /sec: 0.000 vs 13.226 \\
Page-faults                 & 255,776       & 1,162,922     & /sec: 81.944 K vs 4.683 K \\
Cycles                       & 10,088,951,544  & 142,203,906,504 & 3.232 GHz vs 0.573 GHz (util \%) \\
Instructions                 & 15,503,518,578  & 66,061,078,424  & insn per cycle: 1.54 vs 0.46 \\
Branches                     & 3,577,609,384   & 15,816,540,860  & 1.146 G/sec vs 63.698 M/sec \\
Branch-misses                & 76,194,286      & 1,688,310,619   & 2.13\% vs 10.67\% of branches \\
Cache-references             & 1,603,332,840    & 11,744,854,297  & M/sec: 513.667 vs 47.300 \\
Cache-misses                 & 465,208,011      & 5,075,635,536   & 29.02\% vs 43.22\% of cache refs \\
L1-dcache-loads              & 5,212,566,506    & 31,775,281,755  & G/sec: 1.670 vs 0.128 \\
L1-dcache-load-misses        & 593,447,526      & 2,528,078,517   & 11.38\% vs 7.96\% of L1 accesses \\
\hline
\end{tabular}
\end{table}

For testcase \texttt{17.gr}, parallel execution of \texttt{fpt\_subexpo} was much slower than the sequential baseline. The \textbf{task-clock} increased from 3.1~s to 248.3~s, a \textbf{~79x slowdown}. Despite using an average of 2.4~CPUs, the effective \textbf{CPU frequency} dropped from 3.23~GHz to 0.57~GHz, and IPC decreased from 1.54 to 0.46, indicating heavy pipeline stalls and poor per-core efficiency.

Memory performance worsened as well, with \textbf{cache-miss rate} rising from 29.0\% to 43.2\% and \textbf{branch-miss ratio} from 2.13\% to 10.67\%, signaling severe contention.

This behavior is due to the very small values of $|M_t|$ for all $t \in \{1, 2, \ldots, 2n\}$, making parallelization overhead dominate any potential speedup. See \autoref{subexponential:table_17.gr}.

\section{Conclusions}

In this work, we implemented three exact and FPT algorithms for the One-Sided Crossing Minimization problem: the FPT algorithms of Dujmović et al.\ and Kobayashi–Tamaki, as well as a dynamic programming baseline and explored their parallelization options. To support experimentation, we generated test data and made our implementations publicly available \cite{popabogdannnnParallelOSCM}.

Our study highlights the importance of bridging theoretical advances in OSCM algorithms with practical implementations, showing that parallel computing can significantly improve 
their applicability. The results are promising and indicate that further investigation of parallel and distributed strategies for OSCM is both feasible and worthwhile.

\bibliography{lipics-v2021-sample-article}

\appendix

% \section{Implementations}

\section{Bitmask DP Implementation}\label{appendix:bitmask-template}

\begin{algorithm}[H] 
\caption{Computing the optimal number of crossings and reconstructing the solution}
\label{alg:loop}
\scriptsize
\begin{algorithmic}[1]
\Require{$G = (A, B, E)$, $A$ is $\{0, \dots, n - 1\}$, $B$ is $\{0, \dots\ m - 1\}$} 
\Ensure{$P$, optimal permutation of $A$.}
\Statex
\State {Compute $C_{xy}$ for every $x, y \in A$.}
\State {$dp_0 \gets 0$}
\For {$X \gets 1$ to $2^n - 1$}
    \State{$dp_X \gets \infty$}
    \For {$v \gets 0$ to $n - 1$}
        \If{bit $v$ is set in $X$ \textbf{and} $dp_X > dp_{X \oplus 2^v} + F(X \oplus 2^v, v)$}
            \State {$dp_X \gets dp_{X \oplus 2^v} + F(X \oplus 2^v, v)$}
        \EndIf
    \EndFor
\EndFor
\State {$P \gets []$, $X \gets 2^n - 1$}
\While {$X \neq 0$}
    \For {$i \gets 0$ to $n - 1$}
        \If{bit $i$ is set in $X$ \textbf{and} $dp_X = dp_{X \oplus 2^i} + F(X \oplus 2^i, i)$}
            \State{$v \gets i$}
        \EndIf
    \EndFor
    \State {$P \gets [v] + P$} 
    \State {$X \gets X \oplus 2^v$}
\EndWhile
\end{algorithmic}
\end{algorithm}

We can view subsets $X$ of $A$ as binary numbers with $n$ digits, then vertex $v \in X$ if bit $v$ is set in $X$. Recurrence depends on numbers that have one less bit set than $X$, which means that we can compute the array $dp$ in increasing order of $X$. \textbf{Algorithm 1} can run in $O(2^nn^2 + n^2m)$ or $O(2^nn + n^2m)$ depending on the implementation of the $F$ function.

\subsection{Sequential Implementation}\label{appendix:bitmask-sequential}

\begin{algorithm}[H] 
\caption{Calculating the $dp$ array state by state (sequential)}
\label{alg:loop}
\scriptsize
\begin{algorithmic}[1]
\Require{$G = (A, B, E)$, $A$ is $\{0, \dots, n - 1\}$, $B$ is $\{0, \dots\ m - 1\}$} 
\Ensure{$P$, optimal permutation of $A$.}
\Statex
\State {Compute $C_{xy}$ for every $x, y \in A$.}
\State {$dp_0 \gets 0$}
\For {$i \gets 1$ to $n$}
    \For {$k \gets 1$ to ${n \choose i}$}
        \State {$X \gets get\_kth\_state(n, i, k)$}
        \State{$dp_X \gets \infty$}
        \For {$v \gets 0$ to $n - 1$}
            \If{bit $v$ is set in $X$ \textbf{and} $dp_X > dp_{X \oplus 2^v} + F(X \oplus 2^v, v)$}
                \State {$dp_X \gets dp_{X \oplus 2^v} + F(X \oplus 2^v, v)$}
            \EndIf
        \EndFor
    \EndFor
\EndFor
\State {$P \gets []$, $X \gets 2^n - 1$}
\While {$X \neq 0$}
    \For {$i \gets 0$ to $n - 1$}
        \If{bit $i$ is set in $X$ \textbf{and} $dp_X = dp_{X \oplus 2^i} + F(X \oplus 2^i, i)$}
            \State{$v \gets i$}
        \EndIf
    \EndFor
    \State {$P \gets [v] + P$} 
    \State {$X \gets X \oplus 2^v$}
\EndWhile
\end{algorithmic}
\end{algorithm}

Layer $i$ has $n \choose i$ states. We loop through all those states with the help of the function $get\_kth\_state(n, i, k)$ which returns the $k^{th}$ state in $i^{th}$ layer in $O(n)$ time.

\begin{algorithm}[H] 
\caption{$O(n)$ time algorithm for finding the $k^{th}$ state in the $i^{th}$ layer.}
\label{alg:loop}
\scriptsize
\begin{algorithmic}[1]
\Require{$n$, $i$, $k$} 
\Ensure{The $k^{th}$ state in the $i^{th}$ layer.} 
\Statex
\Function{get\_kth\_state}{$n$, $i$, $k$}
    \State {$X \gets 0$}
    \State {$v \gets n - 1$}
    \While{$v >= 0$ \textbf{and} $i > 0$}
        \If{${v \choose i} < k$}
            \State{$k \gets k - {v \choose i}$}
            \State{$i \gets i - 1$}
            \State{$X \gets X + 2^v$}
        \EndIf
        \State{$v \gets v - 1$}
    \EndWhile
    \State \Return {$X$}
\EndFunction
\end{algorithmic}
\end{algorithm}

\textbf{Algorithm 3} constructs the $k^{th}$ state bit by bit, from the most significant to the least significant. For each bit, the algorithm checks if the $k^{th}$ state has it set or not. 
In \textbf{Algorithm 2}, each state is visited exactly one time, computing each state takes $O(n)$ time. For each state we run the same for loop as in \textbf{Algorithm 1}, thus these two algorithms have the same running time.

\subsection{Parallel Implementation}\label{appendix:bitmask-parallel}

\begin{algorithm}[H] 
\caption{Calculating the $dp$ array layer by layer (parallel)}
\label{alg:loop}
\scriptsize
\begin{algorithmic}[1]
\Require{$G = (A, B, E)$, $A$ is $\{0, \dots, n - 1\}$, $B$ is $\{0, \dots\ m - 1\}$} 
\Ensure{$P$, optimal permutation of $A$.}
\Statex
\State {Compute $C_{xy}$ for every $x, y \in A$.}
\State {$dp_0 \gets 0$}
\For {$i \gets 1$ to $n$}
    \For {$j \gets 0$ to $PROC - 1$}
        \If{$j < {n \choose i} \mod PROC$}
            \State{$work_j \gets \ceil{\frac{{n \choose i}}{PROC}}$}
        \Else
            \State{$work_j \gets \floor{\frac{{n \choose i}}{PROC}}$}
        \EndIf
    \EndFor
    \State{$last \gets 0$}
    \For {$j \gets 0$ to $PROC - 1$ \textbf{in parallel}}
        \State{$compute\_range(n, i, last + 1, last + work_j)$}
        \State{$last \gets last + work_j$}
    \EndFor
\EndFor
\State {$P \gets []$, $X \gets 2^n - 1$}
\While {$X \neq 0$}
    \For {$i \gets 0$ to $n - 1$}
        \If{bit $i$ is set in $X$ \textbf{and} $dp_X = dp_{X \oplus 2^i} + F(X \oplus 2^i, i)$}
            \State{$v \gets i$}
        \EndIf
    \EndFor
    \State {$P \gets [v] + P$} 
    \State {$X \gets X \oplus 2^v$}
\EndWhile
\end{algorithmic}
\end{algorithm}

The $compute\_range$ procedure computes the $dp$ values of $[last + 1, last + work_j]$ states in the $i^{th}$ layer.

\begin{algorithm}[H] 
\caption{Implementation of $compute\_range$ procedure. Computes $dp_X$, for every $X$ in range $[a, b]$ on the $i^{th}$ layer.}
\label{alg:loop}
\scriptsize
\begin{algorithmic}[1]
\Require{$n, i, a, b$} 
\Statex
\Function{compute\_range}{$n$, $i$, $a$, $b$}
   \For {$k \gets a$ to $b$}
        \State {$X \gets get\_kth\_state(n, i, k)$}
        \State{$dp_X \gets \infty$}
        \For {$v \gets 0$ to $n - 1$}
            \If{bit $v$ is set in $X$ \textbf{and} $dp_X > dp_{X \oplus 2^v} + F(X \oplus 2^v, v)$}
                \State {$dp_X \gets dp_{X \oplus 2^v} + F(X \oplus 2^v, v)$}
            \EndIf
        \EndFor
    \EndFor
\EndFunction
\end{algorithmic}
\end{algorithm}

\subsection{Optimizing the $F$ Function}

\subsubsection{Fast Bitmask DP}\label{appendix:fast-bitmask}

\begin{algorithm}[H] 
\caption{Precomputation of the $F$ function, sequential}
\label{alg:loop}
\scriptsize
\begin{algorithmic}[1]
\Require{$n, C$}
\Ensure{$F$}
\Statex
\Function{precompute\_F}{$n$, $C$}
    \For {$i \gets 0$ to $n - 1$}
        \For {$conf \gets 1$ to $2^n - 1$}
            \State{$v \gets \floor{{\log_2 conf}}$}
            \State{$F_{conf,i} \gets F_{conf \oplus 2^v, i} + C_{vi}$}
        \EndFor
    \EndFor
    \State \Return $F$
\EndFunction
\end{algorithmic}
\end{algorithm}

\subsubsection{Low-memory Fast Bitmask DP}\label{appendix:low-memory-bitmask}

\begin{algorithm}[H] 
\caption{Precomputation of the $F_1$ and $F_2$}
\label{alg:loop}
\scriptsize
\begin{algorithmic}[1]
\Require{$n, C$}
\Ensure{$F_1$, $F_2$}
\Statex
\Function{precompute\_half\_F}{$n$, $delta$, $s$, $C$}
    \For {$i \gets 0$ to $n - 1$}
        \For {$conf \gets 1$ to $2^s - 1$}
            \State{$v \gets \floor{{\log_2 conf}}$}
            \State{$f_{conf,i} \gets f_{conf \oplus 2^v, i} + C_{v + delta,i}$}
        \EndFor
    \EndFor
    \State \Return $f$
\EndFunction
\State{$F_1 \gets precompute\_half\_F(n, 0, \floor{\frac{n}{2}}, C)$}
\State{$F_2 \gets precompute\_half\_F(n, \floor{\frac{n}{2}}, n - \floor{\frac{n}{2}}, C)$}
\end{algorithmic}
\end{algorithm}

\section{Golden Ratio FPT Implementation}\label{appendix:golden-ratio-fpt}

We will first compute the starting instance $I_0 = (D^0, B^0)$. As proven in \cite{golden_ratio}, for every pair $x, y$ where $C_{xy} = 0$, we know that in the optimal solution $x$ must appear before $y$, so we set $D^0_{xy} = 1$. Note that $D^0$ is transitively closed.

For our implementation we chose to represent instances as objects of class $Instance$ which store the array $D$ and integer $B$ and has several methods:

\begin{algorithm}[H] 
\caption{find\_choice method: returns a \textit{choice pair}}
\label{alg:loop}
\scriptsize
\begin{algorithmic}[1]
\Ensure{A \textit{choice pair}}
\Statex
\Function{find\_choice}{}
    \For {$i \gets 0$ to $n - 1$}
        \For {$j \gets i + 1$ to $n - 1$}
            \If{$D_{ij} = 0$ \textbf{and} $D_{ji} = 0$ \textbf{and} $C_{ij} \neq C_{ji}$}
                \State \Return{$(i, j)$}
            \EndIf
        \EndFor 
    \EndFor
    \State \Return $(-1, -1)$
\EndFunction
\end{algorithmic}
\end{algorithm}

\begin{algorithm}[H] 
\caption{commit method: returns an instance where $D_{ab} = 1$ that is also transitively closed in $O(n^2)$}
\label{alg:loop}
\scriptsize
\begin{algorithmic}[1]
\Ensure{Instance where $D_{ab} = 1$, $D$ transitively closed}
\Statex
\Function{commit}{$a, b$}
    \State {$ret \gets$ \textbf{this}}
    \State {$ret.D_{ab} = 1$}
    \State {$ret.B \gets ret.B - C'_{ab}$}
    \For {$i \gets 0$ to $n - 1$}
        \For {$j \gets 0$ to $n - 1$}
            \If{$ret.D_{ia} = 1$ \textbf{and} $D_{bj} = 1$ \textbf{and} $ret.D_{ij} = 0$}
                \State{$ret.D_{ij} \gets 1$}
                \State{$ret.B \gets ret.B - C'_{ij}$}
            \EndIf
        \EndFor 
    \EndFor
    \State \Return $ret$
\EndFunction
\end{algorithmic}
\end{algorithm}

We create the instance to be returned starting from the object that calls the method. Then we add the new edge. Then for every edge $(i, a)$ and every edge $(b, j)$ we know that there needs to be an edge $(i, j)$ since the resulting $D$ must be transitively closed.

We use a classical \textit{DFS} algorithm to find the topological sorting of $D$. For every vertex $x$, we add its neighbours first in the topological sorting, then we add $x$. The resulting array is then reversed, to give us the actual topological sorting of $D$.

\subsection{Transitive closure optimization}\label{appendix:golden-ratio-tr-closure}

\begin{algorithm}[H] 
\caption{transitive\_closure method: computes and updates $D$ according to its transitive closure in $O(\frac{n^3}{W})$}
\label{alg:loop}
\scriptsize
\begin{algorithmic}[1]
\Ensure{$D$ is transitively closed} 
\Statex
\Function{transitive\_closure}{}
    \State{$srt \gets this.top\_sort()$}
    \For{$i \gets n - 1$ to $0$}
        \For{$j \gets i + 1$ to $n - 1$}
            \If{$D_{srt_i, srt_j} = 1$}
                \State{$D_{srt_i} \gets D_{srt_i} | D_{srt_j}$}
            \EndIf
        \EndFor
    \EndFor
\EndFunction
\end{algorithmic}
\end{algorithm}

The $|$ operation represents the logic \textbf{OR} operation. Since array $D$ is of boolean type, we can group the bits into buckets of length $W$ and use hardware acceleration to speed up our computation. 

\subsection{Search function}\label{appendix:golden-ratio-search-function}

\begin{algorithm}[H] 
\caption{solve function: searches through all instances which add at most $k$ crossings to $LB$}
\label{alg:loop}
\scriptsize
\begin{algorithmic}[1]
\Require{$k$}
\Ensure{The optimal permutation, or empty if a solution couldn't be found with crossing number $\leq k$.} 
\Statex

\Function{search}{I}
    \If{$I.B < 0$}
        \State \Return
    \EndIf
    \State{$(a, b) \gets I.find\_choice()$}

    \If{$a = -1$}
        \If{$sol.B < I.B$}
            \State{$sol.B \gets I.B$}
        \EndIf
        \State \Return
    \EndIf

    \State{$search(I.commit(a, b))$}
    \State{$search(I.commit(b, a))$}
\EndFunction

\Function{solve}{k}
    \State{$I \gets Instance(\emptyset, k)$}
    \State{$I.D_{ij} \gets 1 \forall (i, j) C_{ij} = 0$ \textbf{and} $C_{ji} > 0$}
    \State{$I.D_{ij} \gets 1 \forall (i, j) C'_{ji} > k$}
    \State{$I.transitive\_closure()$}
    \State{$sol \gets (\emptyset, -\infty)$}
    \State{$search(I)$}
    \If{$sol.B < 0$}
        \State \Return $[]$
    \Else
        \State \Return $sol.top\_sort()$
    \EndIf
\EndFunction
\end{algorithmic}
\end{algorithm}

\subsection{Parallel implementation}\label{appendix:golden-ratio-parralel}

The methods of the ThreadPool class are presented below:

\begin{algorithm}[H] 
\caption{push method: pushes an instance to be explored on the stack}
\label{alg:loop}
\scriptsize
\begin{algorithmic}[1]
\Require{Instance $I$}
\Ensure{Instance $I$ is pushed in $S$} 
\Statex

\Function{push}{$I$}
    \State{$acquire(mtx)$}
    \State{$S.push(I)$}
    \State{$release(mtx)$}
    \State{$cv.notify\_one()$}
\EndFunction

\end{algorithmic}
\end{algorithm}

The instance $I$ is pushed onto the stack and then $cv$ notifies one thread that it can stop sleeping and start exploring the new instance that was recently pushed.

\begin{algorithm}[H] 
\caption{push method: pushes an instance to be explored on the stack}
\label{alg:loop}
\scriptsize
\begin{algorithmic}[1]
\Require{Instance $I$}
\Ensure{Instance $I$ is pushed in $S$} 
\Statex

\Function{push}{$I$}
    \State{$acquire(mtx)$}
    \State{$S.push(I)$}
    \State{$release(mtx)$}
    \State{$cv.notify\_one()$}
\EndFunction

\end{algorithmic}
\end{algorithm}

\section{Subexponential FPT Implementation}\label{appendix:subexponential-fpt}

\begin{algorithm}[H] 
\caption{Building the interval system $J$ in $O(E)$ time}
\label{alg:loop}
\scriptsize
\begin{algorithmic}[1]
\Require{Interval system $I$}
\Ensure{Interval system $J$} 
\Statex

\State{Let $s$ be a $m \times 0$ array of pairs}
\For{$a \gets 0$ to $n - 1$}
    \State{$s_{l_i} \gets s_{l_i} + [(a, 0)]$ }
    \State{$s_{r_i} \gets s_{r_i} + [(a, 1)]$ }
\EndFor

\State{$P \gets []$}
\For{$b \gets 0$ to $m - 1$}
    \State{Append to $P$ elements $(a, t)$ from $s_b$ which have $d_a > 1$ and $t = 1$}
    \State{Append to $P$ elements $(a, t)$ from $s_b$ which have $d_a = 1$}
    \State{Append to $P$ elements $(a, t)$ from $s_b$ which have $d_a > 1$ and $t = 0$}
\EndFor

\For{$i \gets 0$ to $2n - 1$}
    \If{$P_i.t = 0$}
        \State{$x_{P_i.a} \gets i$}
    \Else
        \State{$y_{P_i.a} \gets i$}
    \EndIf
\EndFor

\State \Return $J$
\end{algorithmic}
\end{algorithm}

\begin{algorithm}[H] 
\caption{Computing the $dp$ array in $O(k2^{\sqrt{2k}})$}
\label{alg:loop}
\scriptsize
\begin{algorithmic}[1]
\Require{Interval system $I$}
\Ensure{$dp_A$} 
\Statex

\State{Let $event$ be an array of $2n$ pairs}.
\For{$i \gets 0$ to $n - 1$}
    \State{$event_{x_i} \gets (i, 1)$} 
    \State{$event_{y_i} \gets (i, -1)$} 
\EndFor

\State{$L_0 \gets \emptyset$}
\State{$M_0 \gets \emptyset$}
\For{$t \gets 1$ to $2n$}
    \State{$M_t \gets M_{t - 1}$}
    \State{$L_t \gets L_{t - 1}$}
    \If{$event_t.second = 1$}
        \State{$M_t \gets M_t \bigcup \{event_t.first\}$}
    \Else
        \State{$M_t \gets M_t \setminus \{event_t.first\}$}
        \State{$L_t \gets L_t \bigcup \{event_t.first\}$}
    \EndIf
    \For{$S \subseteq M_t$}
        \For{$v \in S$}
            \If{$dp_{L_t \bigcup S} > dp_{L_t \bigcup S \setminus \{v\}} + F(L_t \bigcup S \setminus \{v\}, v)$}
                \State{$dp_{L_t \bigcup S} \gets dp_{L_t \bigcup S \setminus \{v\}} + F(L_t \bigcup S \setminus \{v\}, v)$}
            \EndIf
        \EndFor
    \EndFor
\EndFor
\end{algorithmic}
\end{algorithm}

\end{document}